\documentclass[twoside]{article}
\usepackage{Style.V_7}

\title{Relic Neutrinos Probing Small Scale Primordial Non-Gaussianity}
\paperidentifier{KEK-TH-2875, KEK-Cosmo-0435. Published in \textit{Physics of the Dark Universe} [10.1016/j.dark.2026.102517].}
\paperauthor[1]{Suraj Gavhale}
\paperauthor[2,3]{Shin'ichi Nojiri}
\paperauthor[4,5,6]{Sergei D. Odintsov}
\paperauthor[7]{Oem Trivedi}

\paperaffiliation[1]{Virtual Institute of Astroparticle Physics,
Paris 75018, France}
\paperaffiliation[2]{KEK Theory Center, Institute of Particle and Nuclear Studies, High Energy Accelerator Research Organization (KEK), Tsukuba, Ibaraki 305-0801, Japan}
\paperaffiliation[3]{Kobayashi-Maskawa Institute for the Origin of Particles and the Universe, Nagoya University, Nagoya 464-8602, Japan}
\paperaffiliation[4]{ICREA, Passeig Llu\'is Companys 23, 08010 Barcelona, Spain}
\paperaffiliation[5]{Institute of Space Sciences (ICE-CSIC) C. Can Magrans s/n, Barcelona 08193, Spain}
\paperaffiliation[6]{Institut d'Estudis Espacials de Catalunya (IEEC),
Edifici RDIT, Castelldefels 08860, Spain}
\paperaffiliation[7]{Department of Physics and Astronomy, Vanderbilt University, Nashville 37235, USA}

\newcommand{\R}{\mathcal R}
\newcommand{\PR}{\mathcal P_{\R}}
\newcommand{\BR}{B_{\R}}
\newcommand{\avg}[1]{\left\langle #1\right\rangle}
\newcommand{\dd}{\mathrm d}
\newcommand{\e}{\mathrm e}
\newcommand{\cO}{\mathcal O}
\newcommand{\Mpc}{\mathrm{Mpc}}
\newcommand{\Feff}{F_{\rm eff}}
\newcommand{\Deff}{\mathcal D_{\rm eff}}
\newcommand{\Veff}{\mathcal V_{\rm eff}}
\newcommand{\Wthree}{W^{\rm AR}_{3,M22}}
\DeclareMathOperator{\cyc}{cyc}

\begin{document}
\maketitle

\begin{abstract}
Relic neutrinos retain a record of electromagnetic entropy production after neutrino decoupling, probing primordial modes erased from the primary CMB. We derive the cubic photon number source from acoustic damping and its leading folded sensitive bispectrum window, with neutrino decoupling, thermalization, and acoustic coherence combined in a single response kernel. For excited initial states, this kernel filters the physical folded profile and yields a sharpness dependent transmission condition, directly connecting small scale primordial non-Gaussianity to relic neutrino capture rates.
\end{abstract}

\vfill
\paragraph{Acknowledgment.} The work of OT is supported in part by the Vanderbilt Discovery Doctoral Fellowship.

\newpage

\section{Introduction}
The primary cosmic microwave background (CMB) provides one of the most precise observational windows onto the physics of the early universe, but it offers only a partial record of the primordial perturbation spectrum. On sufficiently small comoving scales, photon diffusion efficiently erases temperature anisotropies before they can survive to recombination, which ends up removing from the primary CMB information about a large range of short wavelength primordial modes, which is relevant both for cosmology and extended theories of gravity \cite{Silk1968,Chluba2012,Khatri2012mix,SunyaevKhatri2013,HuScottSilk1994,ChlubaErickcekBenDayan2012,JeongPradlerChlubaKamionkowski2014,mod1nojiri2017modified,mod2nojiri2007introduction,mod3odintsov2007modified,mod4nojiri2004modified,mod5nojiri2003modified,mod6capozziello2011extended}. This erasure, however, does not mean that the corresponding perturbations leave no physical trace and the acoustic energy stored in these modes is dissipated into the photon-baryon plasma, producing heat and modifying its thermodynamic state. During epochs in which photon number changing processes, most importantly double Compton scattering and bremsstrahlung, remain efficient, the plasma can restore an approximately blackbody spectrum after such energy injection. The restoration of a blackbody then requires the production of additional photons, so that the dissipation of primordial acoustic structure is accompanied by an increase in the comoving photon number and electromagnetic entropy \cite{ChlubaSunyaev2012,KhatriSunyaev2012,Chluba2013Green,PajerZaldarriaga2013,KhatriSunyaevBlackbody2012,ChlubaJeong2014,Tashiro2014}. Consequently, even primordial modes that are absent from the observed CMB anisotropy spectrum can in principle leave an indirect thermodynamic imprint. This observation forms the basis of a broader program in which the thermal history of the photon plasma is used as a probe of primordial fluctuations on scales inaccessible to conventional anisotropy measurements.

Relic neutrinos provide an especially interesting witness to this otherwise hidden part of the thermal history. Once neutrinos decouple from the electromagnetic plasma, entropy subsequently generated in the photon-electron sector is no longer efficiently shared with them. As a result, post decoupling electromagnetic entropy production changes the relation between the neutrino and photon temperatures, reducing $T_\nu/T_\gamma$ relative to the value that would have been obtained in the absence of the additional photon production \cite{Mangano2005,AkitaYamaguchi2020,Piccoli2026,deSalasPastor2016,FrousteyPitrouVolpe2020,BennettBuldgenDrewesWong2020,HannestadMadsen1995}. Since the relic neutrino number density is directly tied to the neutrino temperature, this effect can in turn modify the normalization of the expected event rate in future cosmic neutrino capture experiments \cite{Cocco2007,Betts2013,LongLunardiniSabancilar2014,LazauskasVogelVolpe2008,BettiPTOLEMY2019,FaesslerHodakKovalenkoSimkovic2017}. Recent work by Piccoli, Vagnozzi and Silk \cite{Piccoli2026} exploits precisely this connection, showing that relic neutrino observables can become sensitive to primordial power on scales extending up to approximately $k_D(a_{\rm dec})\simeq 3\times10^5\,\Mpc^{-1}$, far beyond the range preserved in the primary CMB. This opens a particularly intriguing possibility, wherein the same entropy production mechanism may carry not only information about the primordial two-point function, but also about higher order primordial statistics. Spectral distortion observables have already been recognized as probes of small scale primordial non-Gaussianity \cite{PajerZaldarriaga2012,Ota2019ThirdOrder,GancKomatsu2012,EmamiDimastrogiovanniChlubaKamionkowski2015,RavenniLiguoriBartoloShiraishi2017,CabassPajerVanDerWoude2018,pl1planck2020planck,pl2planck2016planck,pl3ade2013planck,pl4akrami2018planck,pl5planck2020planck}, but the relic neutrino problem is structurally different because the relevant observable is the photon number produced by dissipation. A local temperature monopole by itself remains a blackbody and hence cannot be dissipated by Thomson scattering. The physical source must instead be evaluated in the baryon rest frame and must retain the anisotropic component on which the collision term acts \cite{Khatri2012mix,Sharma2024}. Moreover, dissipation occurring while neutrinos remain thermally coupled cannot contribute to the eventual neutrino-photon temperature mismatch. These ingredients mean that the anisotropy structure, the thermalization history, neutrino decoupling and acoustic phase coherence have to be treated together rather than as independent multiplicative corrections.

The central question we would want to then tackle in this work is whether this relic neutrino window can be extended from a probe of small scale primordial power to a controlled probe of small scale primordial non-Gaussianity. If so, what part of the primordial bispectrum can actually survive the combined thermal and acoustic filtering of the early plasma? Building on the thermodynamic connection emphasized in \cite{Piccoli2026}, we derive the cubic photon number source generated by acoustic damping and isolate the leading contribution in the tight coupling regime. Particular care is required because the observable is not simply proportional to the local amount of dissipated energy, but instead the photon number source depends on the monopole anisotropy structure of the radiation field, while the relic neutrino sensitivity further selects only the portion of that dissipation occurring after neutrino decoupling and before photon number changing reactions become ineffective. We hence construct a single response kernel which incorporates neutrino decoupling, blackbody restoration, diffusion damping and acoustic coherence within the same time integral. This formulation reveals that the response is especially sensitive to folded and near folded momentum configurations, for which the acoustic phases can become stationary. We then ask whether such a response can be realized by a physical primordial source rather than by an abstract folded template.

Two big implications can arise from this construction, with the first being that relic neutrino measurements can in principle carry information not only about the amplitude of small scale primordial fluctuations but also about their higher order momentum space structure, providing access to information that has been erased from the primary photon anisotropies. Second, the response is sensitive to the physical history through which the primordial signal is transmitted:, thereby changing the expansion law before BBN alters the relation between conformal time and the diffusion scale, modifies the acoustic kernel, and can shift both neutrino decoupling and the epoch at which photon number changing processes freeze out. The same observable can therefore become simultaneously sensitive to primordial statistics, early universe thermal physics, and non standard pre-BBN expansion. \cite{AllahverdiEtAl2021} The remainder of the paper is organized as follows. In \cref{section 2}, we derive the photon number source through cubic order, identify the leading monopole modulated collision channel in the tight coupling expansion and construct the corresponding primordial bispectrum response kernel, including the combined effects of neutrino decoupling, blackbody restoration, and acoustic coherence. In \cref{section 3}, we apply this response to a physical excited initial state, derive the finite initial time folded profile and its transmission through the acoustic kernel, discuss the resulting sharpness condition together with EFT and backreaction constraints.  We summarize the principal results and discuss their broader implications in \cref{conclusions} (the conclusion), while \cref{appendix} (from the appendix) collects some technical material underlying the main derivations.

\section{Photon number and the cubic collision source}
\label{section 2}

We write the direction dependent photon temperature as $T_\gamma(\bm x,\hat{\bm n},\eta)=\bar T_\gamma(\eta)[1+\Theta(\bm x,\hat{\bm n},\eta)]$ and work in the baryon rest frame.  Here $\eta$ denotes conformal time, $a(\eta)$ the scale factor, $\hat{\mathbf n}$ the photon propagation direction and $\langle\cdots\rangle_{\hat{\mathbf n}}$ an angular average over $\hat{\mathbf n}$.  A prime denotes $d/d\eta$.  We use $\mathcal H\equiv a'/a$ for the conformal Hubble rate and reserve $H\equiv a'/a^2$ for the Hubble rate. We separate the perturbation into its local monopole and anisotropic parts $\Theta=M+B$ with $\avg{B}_{\hat{\mathbf{n}}}=0$.  The energy in excess of a blackbody carrying the same photon number is
\begin{equation}
 \frac{\rho_{\rm ex}}{\bar\rho_\gamma}=\avg{(1+\Theta)^4}_{\hat{\mathbf{n}}}-\avg{(1+\Theta)^3}_{\hat{\mathbf{n}}}^{4/3}=2\avg{B^2}+4M\avg{B^2}+\frac83\avg{B^3}+\cO(\Theta^4).
 \label{eq:excess}
\end{equation}
Every term contains at least two anisotropic factors.  A monopole only changes the temperature of a blackbody, there is nothing for Thomson scattering to mix away.  Complete blackbody restoration conserves the total energy while filling the associated photon number deficit. Since $n_{\rm bb}\propto\rho_{\rm bb}^{3/4}$, we have
\begin{equation}
 \frac{\Delta n_\gamma}{n_\gamma}=\frac34\frac{\rho_{\rm ex}}{\bar\rho_\gamma}+\cO(\rho_{\rm ex}^2),\qquad \left.\frac{\Delta n_\gamma}{n_\gamma}\right|_{(3)}=3M\avg{B^2}+2\avg{B^3}.
 \label{eq:threequarters}
\end{equation}
The conformal Thomson rate is $\Gamma_T \equiv a n_e \sigma_T=-\tau'>0$ where $n_e$ is the physical free electron number density, $\sigma_T$ is the Thomson cross section, and $\tau$ is the Thomson optical depth. In an arbitrary scalar gauge we define the gauge invariant baryon-frame anisotropy and comoving monopole by
\begin{equation}
  B \equiv \Theta-\Theta_0-\hat{\mathbf n}\!\cdot\!\mathbf v_b
   =\hat{\mathbf n}\!\cdot(\mathbf v_\gamma-\mathbf v_b)
    +\Theta_{\ell\ge2},
\end{equation}
\begin{equation}
  M \equiv \widetilde\Theta_0
   =\Theta_0+\mathcal H v_b .
\end{equation}
Here $\mathbf v_b$ and $\mathbf v_\gamma$ are the baryon and photon peculiar velocities (or their scalar velocity potentials in Fourier space, according to the convention used below). We denote the linear and quadratic Thomson collision images by $C^{(1)}$ and $C^{(2)}$. In the temperature shift sector of the second order collision operator \cite{Pitrou2009,BenekeFidler,Sharma2024,PitrouUzanBernardeau2010}, $C^{(1)}=-B-\Pi^{(1)}_2$ and $C^{(2)}=B^2-\Pi^{(2)}_2$ where $\Pi^{(r)}_2$ is the pure quadrupole polarization collision term at perturbative order $r$. Differentiating \cref{eq:excess} along photon geodesics and using this collision operator, with $Q\equiv\rho_{\rm ex}$ denoting the excess photon energy density, gives
\begin{align}
 \frac{\dd Q}{\dd\eta}=-4\bar\rho_\gamma\Gamma_T\Big[&\avg{BC^{(1)}}+\avg{BC^{(2)}}+2\avg{B^2C^{(1)}}+2\avg{MBC^{(1)}}\Big]+\cO(\Theta^4).
 \label{eq:full-cubic-heating}
\end{align}
This with \cref{eq:threequarters} gives the photon number rate
\begin{align}
 \frac{\dd}{\dd\eta}\ln(n_\gamma a^3)=-3\Gamma_T A_{\rm bb}(\eta)\Big[&\avg{BC^{(1)}}+\avg{BC^{(2)}}+2\avg{B^2C^{(1)}}+2\avg{MBC^{(1)}}\Big].
 \label{eq:full-number-source}
\end{align}
The factor $A_{\rm bb}(\eta)$ is the blackbody restoration efficiency associated with photon producing thermalization processes. It is normalized to unity when double Compton scattering and bremsstrahlung efficiently restore a blackbody and tends to zero once photon number changing reactions have frozen out. 

\Cref{eq:full-number-source} contains every cubic contribution generated within the second order collision or temperature shift construction.  A third order kinetic derivation may also contain frequency dependent structures and collision terms that cannot be inferred from \cref{eq:excess} and this is emphasized in ~\cite{Sharma2024}.  With that, we use the standard \cite{MaBertschinger1995} multipole convention
\begin{equation}
 \Theta(\bm k,\hat{\bm n},\eta)=\sum_{\ell\ge0}(-i)^\ell(2\ell+1)\Theta_\ell(k,\eta)P_\ell(\hat{\bm k}\!\cdot\!\hat{\bm n})\R_{\bm k}.
\end{equation}
Here $P_\ell$ is the Legendre polynomial, $\mathcal R_{\mathbf k}$ is the primordial comoving curvature perturbation, $\Theta_\ell$ are the photon temperature transfer multipoles, and $\Theta^P_\ell$ denotes the corresponding linear polarization multipoles.  We use $R_b\equiv {3\bar\rho_b}/{4\bar\rho_\gamma}$ for the baryon loading parameter.  The symbol $A$ appearing in the order estimates denotes a characteristic linear perturbation amplitude for the triangle under consideration. The baryon-frame dipole, the polarization corrected quadrupole collision image and the higher multipoles then read
\begin{align}
 B_1&=\Theta_1-\frac{k v_b}{3}=\frac{k}{3}(v_\gamma-v_b),&C^{(1)}_1&=-B_1, \label{eq:dipole-slip}\\B_2&=\Theta_2,&C^{(1)}_2&=-\Theta_2+\frac1{10}(\Theta_2+\Theta^P_0+\Theta^P_2)\simeq-\frac34\Theta_2, \label{eq:quad-pol}\\
 B_\ell&=\Theta_\ell,&C^{(1)}_\ell&=-\Theta_\ell,\qquad \ell\ge3. \label{eq:higher-collision}
\end{align} 
The last equality in \cref{eq:quad-pol} uses the leading tight coupling polarization relation $\Theta_2^P+\Theta_0^P\simeq3\Theta_2/2$ \cite{Chluba2012,Sharma2024}.  Let $\epsilon_a=k_a/\Gamma_T$ at the time the triangle is damped and $\epsilon=\max_a\epsilon_a\ll1$.  A regular tight coupling solution \cite{HuSugiyama1996,CLASSII,Sharma2024,CyrRacineSigurdson2011} obeys $M=\cO(A)$, $\Theta_2,\Theta^P_0,\Theta^P_2=\cO(\epsilon A)$ and $\Theta_{\ell}=\cO(\epsilon^{\ell-1}A)$ {($\ell\ge3$)} with $B_1=O(R_b\Theta_2)=O(R_b\epsilon A)$ and $\Theta_3\simeq({3k}/{7\Gamma_T})\Theta_2$.

The second order polarization collision term is regular in the same expansion, $\Pi^{(2)}_2=\cO(B^2)=\cO(\epsilon^2A^2)$.  Substituting these and staying away from a phase or angular zero of the leading integrand gives $\avg{MBC^{(1)}}=\cO(\epsilon^2A^3)$ with $\avg{B^2C^{(1)}}=\cO(\epsilon^3A^3)$ and  $\avg{BC^{(2)}}=\cO(\epsilon^3A^3).$ Thus, within this construction, each of the other two cubic terms is smaller than the monopole modulated term by one power of $\epsilon$.  Inside the leading channel itself, dipole slip contributes at relative order $\cO(R_b^2)$ while the $\ell=3$ term is suppressed by $\cO(\epsilon^2)$ relative to the polarization corrected quadrupole. The counting follows from the fact that every factor of $B$ or $C^{(1)}$ costs one power of $\epsilon$ whereas $M$ costs none. Indeed, it is important not to ask more of this counting than it proves. Where the leading channel cancels (as in the equilateral asymptotic) the $\cO(\epsilon^3A^3)$ terms may determine the residual.  Nor can the above averages constrain new terms that only a third order photon Boltzmann calculation could. We turn the leading collision channel into a primordial bispectrum window.  Our convention is
\begin{equation}
 \avg{\R_{\bm k_1}\R_{\bm k_2}\R_{\bm k_3}}=(2\pi)^3\delta^{(3)}(\bm k_1+\bm k_2+\bm k_3)\BR(k_1,k_2,k_3).
 \label{eq:bispectrum-def}
\end{equation}
We denote by $W^{\rm AR}_{3,M22}$ the cubic acoustic reheating response window of the leading monopole modulated channel, and $M22$ indicates one monopole transfer factor and the leading pair of polarization corrected quadrupole collision factors. The sum over $\mathfrak S_3$ runs over all six permutations of $(i,j,m)$. Projecting the $MBC^{(1)}$ term of \cref{eq:full-number-source} gives
\begin{align}
 \Wthree(k_1,k_2,k_3)=-\int\dd\eta\,\Gamma_T V_\nu(\eta)\sum_{(i,j,m)\in {\mathfrak S_3}}\sum_{\ell\ge1}(-1)^\ell(2\ell+1) \widetilde T_0(k_i)B_\ell(k_j)C^{(1)}_\ell(k_m)P_\ell(\mu_{jm})
 \label{eq:general-window}
\end{align}
where $\mu_{jm}={k_i^2-k_j^2-k_m^2}/{2k_jk_m}$ and $V_\nu(\eta)=G_\nu(\eta)A_{\rm bb}(\eta)$. The response $G_\nu$ vanishes while neutrinos still share the injected entropy and tends to unity after decoupling.  Notice that there is no factor of $1/3!$ and the six assignments are needed to reproduce the coefficient six multiplying $\avg{MBC^{(1)}}$ in the photon number source.  For an isotropic bispectrum, the result becomes
\begin{equation}
 \Delta\ln(n_\gamma a^3)^{(3)}_{M22}=\frac{1}{8\pi^4}\int_\triangle\prod_{i=1}^3\dd k_i\,k_1k_2k_3\, \BR(k_1,k_2,k_3)\Wthree(k_1,k_2,k_3).
\label{eq:entropy-bispectrum}
\end{equation}
Here the integral is over the triangle domain $|k_i-k_j|\le k_m\le k_i+k_j$.

\Cref{eq:general-window} is therefore the leading monopole modulated, linear collision projection and it should not be confused with the cubic window associated with \cref{eq:full-number-source}.  Let $A_\nu=(1+4f_\nu/15)^{-1}$, where $f_\nu\equiv \bar\rho_\nu/(\bar\rho_\gamma+\bar\rho_\nu)$ is the neutrino fraction of the radiation energy density, and $A_\nu$ accounts for the free streaming neutrino suppression of the acoustic amplitude \cite{BashinskySeljak2004}. We take $c_s^2=1/3$ and call
\begin{equation}
  r_s(\eta)\equiv \int_0^\eta c_s(\eta')\,d\eta'
  \simeq c_s\eta
\end{equation}
for the photon sound horizon in the radiation era approximation.  Well inside the horizon during radiation domination, we have
\begin{equation}
     \widetilde T_0(k,\eta)\simeq A_\nu\cos(kr_s)\e^{-k^2/k_D^2},
     \label{eq:tca-transfer0}
\end{equation}
\begin{equation}
 \Theta_2(k,\eta)\simeq \frac{8kc_s}{15\Gamma_T}A_\nu\sin(kr_s)\e^{-k^2/k_D^2}.
 \label{eq:tca-transfer2}
\end{equation}
Using the polarization corrected collision image in \cref{eq:quad-pol} then gives
\begin{align}
 \Wthree\simeq{}&\frac{32A_\nu^3}{45}\int_0^\infty\frac{\dd\eta}{\Gamma_T}V_\nu(\eta)\exp\!\left[-\frac{q}{k_D^2}\right]\sum_{\cyc}k_jk_mP_2(\mu_{jm})\cos(k_ir_s)\sin(k_jr_s)\sin(k_mr_s),
 \label{eq:tca-window}
\end{align}
where $q=k_1^2+k_2^2+k_3^2$. The time integral becomes clear after converting conformal time for the diffusion scale.  For $a(\eta)=H_0\sqrt{\Omega_r}\eta$, $\Gamma_T=\Gamma_*/\eta^2$ and $\dd k_D^{-2}/\dd\eta=8/(45\Gamma_T)$, we have $y\equiv k_D^{-2}={8\eta^3}/{135\Gamma_*}$, $\eta_q=({135\Gamma_*}/{8q})^{1/3}$ and $r_q=c_s\eta_q$. It follows that $u=\eta/\eta_q=(qy)^{1/3}$ and, crucially, that $y=u^3/q$. This motivates a kernel
\begin{equation}
 \Feff(x;q)=3\int_0^\infty\dd u\,u^2\e^{-u^3}V_\nu\!\left(\frac{u^3}{q}\right)\cos(xu).
 \label{eq:Feff-def}
\end{equation}
The argument $u^3/q$ has the dimensions of $y=k_D^{-2}$ while the superficially similar choice $V_\nu(qu^3)$ is dimensionally incorrect.  A product to sum reduction of \cref{eq:tca-window} yields
\begin{align}
 \Wthree=\frac{A_\nu^3}{q}\sum_{\cyc}k_jk_mP_2(\mu_{jm})\big[&-\Feff((k_i-k_j-k_m)r_q;q)\nonumber\\
 &+\Feff((k_i+k_j-k_m)r_q;q)\nonumber\\
 &+\Feff((k_i-k_j+k_m)r_q;q)\nonumber\\
 &-\Feff((k_i+k_j+k_m)r_q;q)\big].
 \label{eq:master-window}
\end{align}
As a useful reference point, complete thermalization with no neutrino gate means $V_\nu=1$ and reduces \cref{eq:Feff-def} to
\begin{equation}
 F(x)=3\int_0^\infty\dd u\,u^2\e^{-u^3}\cos(xu)=\sum_{n=0}^\infty\frac{(-1)^n\Gamma(1+2n/3)}{(2n)!}x^{2n}\sim\frac{360}{x^6}+\cO(x^{-12})\quad(\text{as } x\to\infty)
 \label{eq:F-def}
\end{equation}
and its first zero is $x_0=1.7650100459\ldots$. Consider a nondegenerate folded triangle $k_1=k_2+k_3$ with $k_i r_q\gg1$.  The three stationary weights add to $q/2$ and
\begin{equation}
 \left.\Wthree\right|_{\rm folded}=\frac{A_\nu^3}{2}\Feff(0;q)+\cO[(k_{\min}r_q)^{-6}],
 \label{eq:folded-height-effective}
\end{equation}
\begin{equation}
 \Veff(q)\equiv\Feff(0;q)=q\int_0^\infty\dd y\,\e^{-qy}G_\nu(y)A_{\rm bb}(y).
 \label{eq:Veff}
\end{equation}
With the response functions defined above, we can take instantaneous neutrino decoupling and complete thermalization, $G_\nu(y)=\Theta(y-y_{\rm dec})$ and $A_{\rm bb}=1$, where $y_{\rm dec}\equiv k_D^{-2}(a_{\rm dec})$. Then $\Veff(q)=\e^{-qy_{\rm dec}}$, so that the folded height is
\begin{equation}
 \left.\Wthree\right|_{\rm folded}=\frac{A_\nu^3}{2}\exp\!\left[-\frac{q}{k_D^2(a_{\rm dec})}\right].
 \label{eq:gated-folded-height}
\end{equation}
The value $A_\nu^3/2=0.36655$ is therefore reached deep in the post-decoupling regime, $q\ll k_D^2(a_{\rm dec})$. The size of this effect is easy to see.  For a folded $3{:}2{:}1$ triangle, $q=(14/9)k_1^2$.  Taking $k_D(a_{\rm dec})=3\times10^5\,\Mpc^{-1}$, the sharp gate factors at $k_1=5\times10^4$ and $3\times10^5\,\Mpc^{-1}$ are $0.9577$ and $0.2111$ respectively.  The corresponding ideal folded heights are $0.3510$ and $0.0774$.  In a realistic calculation, a smooth decoupling history simply replaces the step by the physical $G_\nu(y)$ in \cref{eq:Veff}.

To resolve the shape near the boundary, let $\delta=k_2+k_3-k_1\ge0$ and we can define
\begin{equation}
 G(k_1,k_2,k_3)=-k_2k_3P_2(\mu_{23})+k_3k_1P_2(\mu_{31})+k_1k_2P_2(\mu_{12}).
\end{equation}
\begin{equation}
 W^{\rm AR}_{3,M22,{\rm slow}}\simeq\frac{A_\nu^3}{q}G(k_1,k_2,k_3)\Feff(\delta r_q;q)
 \label{eq:near-folded}
\end{equation}
with $G|_{\delta=0}=q/2$. For a band that is uniform over $0\le\delta\le\Delta\delta$, let $X=r_q\Delta\delta$ and we define
\begin{equation}
 \Deff(X;q)=\frac{1}{X\Feff(0;q)}\int_0^X\dd x\,\Feff(x;q).
 \label{eq:Deff}
\end{equation}
The band averaged response is then $\Feff(0;q)\Deff(X;q)$.  For comparison, we define the ungated band average $\mathcal D(X)\equiv X^{-1}\int_0^X F(x)\dd x$.  In general, $\Feff(x;q)\ne\Feff(0;q)F(x)$ and the gated response does not factorize as ${\Deff(X;q)=\mathcal D(X)}$. 

Blackbody restoration, neutrino decoupling and acoustic coherence are different weights in the same integral. The reason a finite width matters can be seen already in the ungated limit.  For $V_\nu=1$, the kernel $F$ is compensated and $\int_0^\infty F(x)\dd x=0$. Its first moment (integral $xF(x)$) is $-\Gamma(1/3)$ and its second moment (integral $x^2F(x)$) is $-3\pi$. For a smooth compactly supported normal profile $\Psi$, we get
\begin{equation}
 \int_0^\infty\dd\delta\,{\Psi(\delta)F(r_q\delta)=-\frac{\Gamma(1/3)}{r_q^2}\Psi'(0)-\frac{3\pi}{2r_q^3}\Psi''(0)+\cO(r_q^{-4})}.
 \label{eq:smooth-asymptotic}
\end{equation}
Here $\Psi(\delta)$ denotes a smooth profile in the direction normal to the folded boundary, with $\delta=k_2+k_3-k_1$. The constant boundary value therefore cancels at order $r_q^{-1}$.  As before, this belongs specifically to the leading $M22$ channel and is not about the sum of every possible third order collision source.

\begin{figure}[t]
 \centering
 \includegraphics[width=\textwidth]{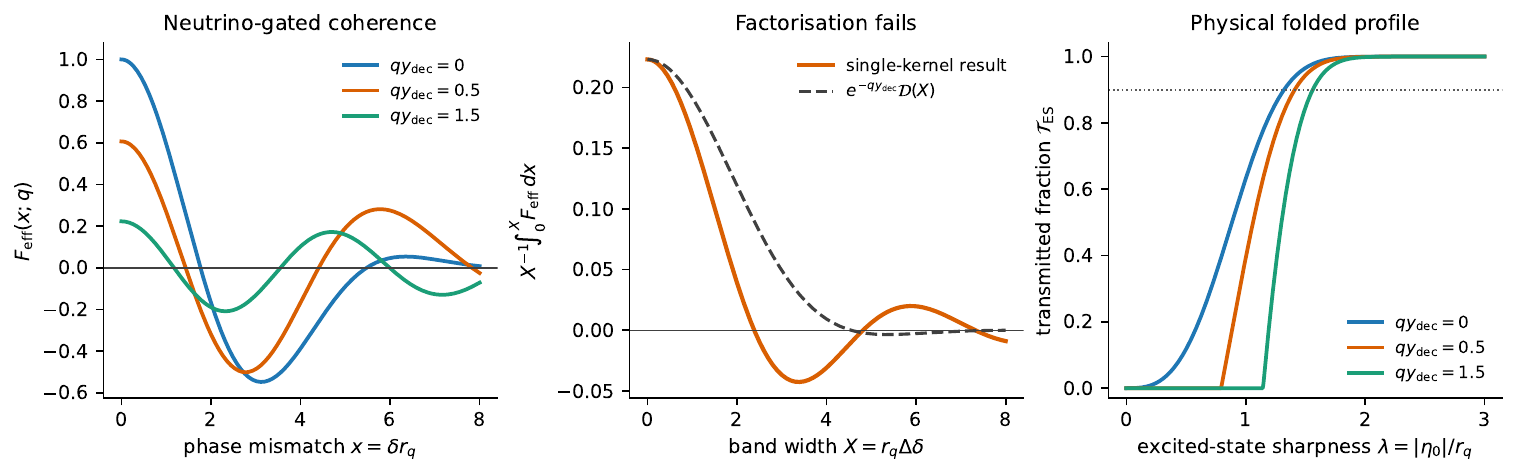}
 \caption{Corrections that cannot be represented by the original product of independent factors. \textit{Left:} the sharp neutrino gate changes both the height and the phase structure of the coherence kernel. \textit{Centre:} for $qy_{\rm dec}=1.5$, the exact band response differs from $\exp(-qy_{\rm dec})\mathcal D(X)$. \textit{Right:} transmission of the physical finite-initial-time sinc profile derived. These are not a CLASS transfer function comparison.}
 \label{fig:kernel-diagnostics}
\end{figure}

\section{A physical excited state bispectrum}
\label{section 3}
The response derived above tells us what the late time plasma accepts, but a phenomenological folded band does not yet tell us what inflation supplies.  In particular, the band normalized parameter $f_{\rm fold}$ is a useful forecast coordinate, not an inflationary model.  We therefore use a finite initial time bispectrum before introducing any effective amplitude.  Excited initial states are natural for this comparison because their bispectra can be enhanced near collinear or folded configurations \cite{Meerburg2009,AgulloParker2011,Ganc2011,AshoorioonShiu2011}.

For a Gaussian Bogoliubov initial state specified at conformal time $\eta_0<0$, the minimally coupled single field calculation of Holman and Tolley gives, to first order in $\beta_k$ \cite{HolmanTolley} as
\begin{equation}
 B_{\R}^{\rm HT}(k_1,k_2,k_3)=-\frac{4H^6}{M_{\rm Pl}^2\dot\phi^2\prod_a(2k_a^3)}\sum_{j=1}^3\frac{k_1^2k_2^2k_3^2}{k_j^2}\operatorname{Re}\!\left[\beta_{k_j}^*\frac{1-\e^{i\widetilde k_j\eta_0}}{\widetilde k_j}\right]
 \label{eq:HT-bispectrum}
\end{equation}
where $\widetilde k_j\equiv k_1+k_2+k_3-2k_j$. Here $H$ is the inflationary Hubble rate, $M_{\rm Pl}$ is the reduced Planck mass, $\dot\phi$ is the cosmic time derivative of the homogeneous inflaton, $\beta_k$ is the Bogoliubov excitation coefficient. The factor $1/\widetilde k_j$ looks singular, but the numerator regulates it.  On the boundary $k_1=k_2+k_3$, $\widetilde k_1=\delta$.  The imaginary part of $\beta_{k_1}$ produces the finite folded profile $\sin(\delta\eta_0)/\delta$, while the real part starts linearly in $\delta$. The physical prediction is obtained by inserting \cref{eq:HT-bispectrum} directly into \cref{eq:entropy-bispectrum} and
\begin{equation}
 S_3^{\rm HT}=\frac{1}{8\pi^4}\int_\triangle\prod_{i=1}^3\dd k_i\,k_1k_2k_3\,B_{\R}^{\rm HT}(k_1,k_2,k_3)\Wthree(k_1,k_2,k_3).
 \label{eq:S3HT-exact}
\end{equation}
Neither $f_{\rm fold}$ nor a top hat width appears in this equation. The normal momentum convolution can be performed analytically.  Let $x=r_q\delta$ and $\lambda=|\eta_0|/r_q$ and normalize the imaginary $\beta$ profile to its boundary value $\varphi_\lambda(x)={\sin(\lambda x)}/{\lambda x}.$ Then
\begin{equation}
 \int_0^\infty\dd x\,F(x)\varphi_\lambda(x)=\frac{\pi}{2\lambda}\left(1-\e^{-\lambda^3}\right).
 \label{eq:HT-convolution}
\end{equation}
Relative to an uncompensated filter with the same stationary height, the exact acoustic transmission is $\mathcal T_{\rm ES}(\lambda)=1-\e^{-\lambda^3}$. For a broad physical feature, $|\eta_0|\ll r_q$, the transmission begins cubically with $\mathcal T_{\rm ES}\sim\lambda^3$.  It reaches $50\%$, $90\%$ and $99\%$ at $\lambda=0.885$, $1.321$ and $1.664$, respectively. Restoring the complete time response gives
\begin{equation}
 \mathcal T_{\rm ES}(\lambda;q)=\frac{\displaystyle\int_0^\lambda\dd u\,3u^2\e^{-u^3}V_\nu(u^3/q)}{\Feff(0;q)}.
 \label{eq:HT-transmission-general}
\end{equation}
For the sharp neutrino gate used above (for \cref{eq:gated-folded-height}), with $s=qy_{\rm dec}$, we have
\begin{equation}
 \mathcal T_{\rm ES}(\lambda;s)=\Theta(\lambda^3-s)\left[1-\e^{-(\lambda^3-s)}\right].
 \label{eq:HT-transmission-gated}
\end{equation}
This gives the selection rule in its most direct form. The feature must be sharp enough to overlap the post-decoupling portion of the damping kernel.  Otherwise the response vanishes, even when the bispectrum is large on the folded boundary.

There is, however, a cost to making the feature narrow.  For a mode excited below the EFT cutoff $M_{\rm EFT}$, $k|\eta_0|\lesssim M_{\rm EFT}/H$.  Achieving $90\%$ acoustic transmission requires ${M_{\rm EFT}}/{H}\gtrsim1.321\,k r_q.$ At the benchmark $k r_q\simeq2.17\times10^3$, this becomes $M_{\rm EFT}/H\gtrsim2.9\times10^3$.  The same excited state is also constrained by backreaction.  The slow roll bound of ~\cite{HolmanTolley} has the parametric form
\begin{equation}
 |\beta_k|\lesssim\sqrt{\epsilon_{\rm sr}\eta_{\rm sr}}\frac{HM_{\rm Pl}}{{M_{\rm EFT}}^2}=\frac{\sqrt{\eta_{\rm sr}}}{\sqrt{8\pi^2\PR(k)}}\left(\frac{H}{{M_{\rm EFT}}}\right)^2.
 \label{eq:beta-bound}
\end{equation}
Here ${M_{\rm EFT}}$ is the EFT cutoff of the excited initial state and $\epsilon_{\rm sr}$ and $\eta_{\rm sr}$ are the slow roll parameters appearing in the backreaction bound. Because the unprojected folded enhancement scales as $|\beta_k|k|\eta_0|$, increasing ${M_{\rm EFT}}/H$ pulls in opposite directions. It sharpens the feature while tightening the allowed occupation number. This tension is familiar in controlled excited state constructions \cite{FlaugerGreenPorto2013,CollinsHolman2005,SchalmShiuVanDerSchaar2004,GreeneSchalmShiuVanDerSchaar2005}.  For modes at the cutoff, the filtered parametric enhancement is bounded by
\begin{equation}
 \frac{\sqrt{\eta_{\rm sr}}}{\sqrt{8\pi^2\PR(k)}}\frac{1}{{M_{\rm EFT}}/H}\left[1-\exp\!\left(-\left(\frac{{M_{\rm EFT}}/H}{kr_q}\right)^3\right)\right],
 \label{eq:coherence-backreaction}
\end{equation}
up to the smooth momentum and slow-roll factors shown in \cref{eq:HT-bispectrum}. The competition inside the brackets is maximized at $({M_{\rm EFT}}/H)/(kr_q)=1.239\ldots$.  For the illustrative choice $\PR=0.06$, $\eta_{\rm sr}=10^{-2}$ and $kr_q=2.17\times10^3$, the bound is of order $1.5\times10^{-5}$.  This is not a universal limit on every excited state model.  Higher derivative interactions carry different powers of ${M_{\rm EFT}}/H$ and demand their own EFT and backreaction analysis.  What the estimate does show is more specific, that the minimal controlled model cannot realize an arbitrarily narrow band with a large $f_{\rm fold}$.

For the narrow band forecast we define $\alpha\equiv \mathcal P_{\mathcal R}(k_*)$ as the approximately constant dimensionless small scale curvature power over the band centered on the characteristic scale $k_*$. The quantities $S_2$ and $S_3$ denote, respectively, the quadratic and cubic contributions to the logarithmic comoving photon number shift used in the one bin forecast.  The symbol $N$ is the expected relic neutrino capture count and $\sigma_\alpha$ is the width of an external Gaussian prior on $\alpha$. The calculation does not make the phenomenological template useless, rather, it tells us how to interpret it. Since $\PR(k)$ is dimensionless, the corresponding dimensional two-point factor is ${(2\pi^2/k^3)\PR(k)}$.  For a boundary localized template, we may therefore define
\begin{align}
 \BR^{\rm fold}=\frac65f_{\rm fold}{(2\pi^2)^2}\left[{\frac{\PR(k_1)\PR(k_2)}{k_1^3k_2^3}}U_\Delta(k_3;k_1,k_2)+2\ {\rm cyclic}\right]
\end{align}
\begin{equation}
 \int_{|k_i-k_j|}^{k_i+k_j}\frac{k_m\dd k_m}{k_ik_j} U_\Delta(k_m;k_i,k_j)=1.
\end{equation}
For a narrow uniform band, the cubic signal is $ S_3=\mathcal C_3\,\Feff(0;q)\Deff(X;q) f_{\rm fold}\alpha^2$ and $\mathcal C_3={9A_\nu^3}/{10}=0.65978.$ The amplitude $f_{\rm fold}$ remains a mathematical coordinate on this family of bands, not a prediction of \cref{eq:HT-bispectrum}. The same gate also enters the quadratic coefficient.  In the deep acoustic regime, we have $S_2=\mathcal C_2(k_*)\alpha$ and $\mathcal C_2(k_*)\simeq({3A_\nu^2}/{2}) \Feff(0;2k_*^2)$. For a Poisson count $N$ and an external Gaussian prior $\sigma_\alpha$, we get
\begin{equation}
 \sigma(f_{\rm fold})= \frac{\sqrt{N^{-1}+\mathcal C_2(k_*)^2\sigma_\alpha^2}}{\mathcal C_3\alpha^2|\Feff(0;q)\Deff(X;q)|}.
 \label{eq:fisher-corrected}
\end{equation}
Without the power prior, the Fisher matrix remains rank one.  In the combined deep post-decoupling, complete thermalization and zero width limit, $N=100$ and $\alpha=0.06$ reproduce the ideal value $\sigma(f_{\rm fold})=42.1$.  At the upper scale of the sharp gate $3{:}2{:}1$ example, the response factor $0.2111$ alone degrades this to approximately $199$. 

The calculation now reaches the point where a numerical test is both possible and necessary.  The appropriate check uses linear transfer functions exported from a modified CLASS hierarchy and second order Boltzmann solvers such as SONG illustrate the level required when polarization is retained \cite{Pettinari2014,FidlerKoyamaPettinari2015}.  For every $k$ and time, the required quantities are $\widetilde\Theta_0, v_\gamma-v_b,\Theta_2,\Theta^P_0,\Theta^P_2, \Theta_\ell\ (3\le\ell\le\ell_{\max}), \Gamma_T$ and $k_D$. Together, these quantities reconstruct \cref{eq:dipole-slip,eq:quad-pol,eq:higher-collision} and can be inserted directly into \cref{eq:general-window}.  A convincing numerical analysis should show $W^{\rm AR}_{3,M\times\mathrm{linear}}/W^{\rm AR}_{3,M22}$ across the folded boundary for equilateral, folded and near folded triangles \footnote{For the present work, we have not generated a CLASS table.}. It should also show the fractional error in the deep window $0.36655$ value, separately displaying the neutrino gate and the signed compensation integral for a smooth folded template.

The kernel above was obtained assuming radiation domination, $a(\eta)\propto\eta$.  Its dependence on the background expansion can be made explicit, which also shows how the relic neutrino response can probe a non-standard pre-BBN expansion history. \cite{DEramoFernandezProfumo2017}  Consider a stage in which $a(\eta)\propto \eta^p$ with ${p>0}$ and assume, over the interval relevant for damping, that the effective Thomson scattering lepton density scales approximately as $n_e\propto a^{-3}$.  The conformal Thomson rate then
obeys $\Gamma_T=a n_e\sigma_T\propto a^{-2}\propto\eta^{-2p}.$ If the slowly varying baryon loading corrections in the diffusion coefficient are neglected at this stage, the tight coupling diffusion gives ${d k_D^{-2}}/{d\eta}\propto \Gamma_T^{-1}\propto \eta^{2p}$ with $k_D^{-2}\propto \eta^{2p+1}.$ It is useful to define $m\equiv 2p+1$ with $q\,y(\eta_q)=1$ and $u\equiv {\eta}/{\eta_q}$ so that $qy=u^m$. The radiation dominated case corresponds to $p=1$ and hence $m=3$. With these definitions, the response kernel generalizes to
\begin{equation}
  F_{\rm eff}^{(p)}(x;q)= m\int_0^\infty du\,u^{m-1}\e^{-u^m}V_\nu\!\left(\frac{u^m}{q}\right)\cos(xu),
  \label{eq:kernel-general-p}
\end{equation}
where $x$ is the acoustic phase mismatch measured in units of $r_q=c_s\eta_q$. \Cref{eq:kernel-general-p} reduces to \cref{eq:Feff-def} for $p=1$.  In the ungated limit $V_\nu=1$, the finite initial time sinc profile can again be convolved analytically, and the acoustic transmission becomes
\begin{equation}
  \mathcal T_{\rm ES}^{(p)}(\lambda)=1-\e^{-\lambda^{\,m}}=1-\e^{-\lambda^{\,2p+1}}.
  \label{eq:transmission-general-p}
\end{equation}
More generally, including neutrino decoupling and blackbody thermalization in the same kernel, we have
\begin{equation}
  \mathcal T_{\rm ES}^{(p)}(\lambda;q)=\frac{\displaystyle\int_0^\lambda du\,m u^{m-1}\e^{-u^m}V_\nu(u^m/q)}{F_{\rm eff}^{(p)}(0;q)}.
  \label{eq:transmission-general-p-gated}
\end{equation}
For an instantaneous neutrino gate $G_\nu(y)=\Theta(y-y_{\rm dec})$ with complete blackbody thermalization, $s\equiv qy_{\rm dec}$ gives
\begin{equation}
  \mathcal T_{\rm ES}^{(p)}(\lambda;s)=\Theta(\lambda^m-s)\left[1-\e^{-(\lambda^m-s)}\right].
\end{equation}
A modified expansion history also changes the neutrino gate itself. The physical Hubble rate $H$ sets the neutrino decoupling epoch through the competition $\Gamma_{\nu e}(T_{\rm dec})\simeq H(T_{\rm dec})$, where $\Gamma_{\nu e}$ is the weak neutrino--electron interaction rate. Changing $H(T)$ therefore shifts both the decoupling temperature and the response $G_\nu$, and can also shift the freeze out of photon number changing processes encoded by $A_{\rm bb}$, in addition to changing the mapping between conformal time and the diffusion scale \cite{KawasakiKohriSugiyama2000}. Consequently the $k$-interval to which the relic neutrino observable is most sensitive can move in a non-standard cosmology. \cite{GelminiGondolo2008} Such a shift can arise, for example, in modified gravity backgrounds, although the response derived here constrains the expansion history itself rather than any specific gravitational model.

\section{Conclusions}
\label{conclusions}
In this work, we have developed a framework through which relic neutrinos can probe primordial non-Gaussianity on scales that are erased from the primary CMB by photon diffusion. Starting from the photon number response to acoustic dissipation, we derived the cubic collision source in the baryon rest frame and identified the leading monopole modulated $M22$ channel responsible for transmitting primordial bispectrum information into the electromagnetic entropy history. We showed that neutrino decoupling, blackbody restoration and acoustic coherence cannot, in general, be treated as independent multiplicative effects. Instead, they all enter through a common response kernel $F_{\rm eff}$ and this construction exposes a pronounced sensitivity to folded and near-folded momentum configurations. For physical excited initial states, we derived the corresponding finite initial time folded profile and its exact acoustic transmission, obtaining a sharpness dependent selection rule, wherein only features sufficiently narrow to overlap the post decoupling portion of the damping kernel are efficiently transmitted to the relic-neutrino observable. We further connected this response to a phenomenological folded amplitude forecast, while making it clear that EFT and backreaction restrictions that prevent arbitrarily sharp and arbitrarily large signals.

The broader implication of these results is that relic neutrino capture experiments could provide access to a qualitatively new window on the primordial Universe, extending quite beyond the comoving scales directly accessible through conventional CMB anisotropies. Instead of merely constraining the amplitude of small scale primordial power, such measurements may retain information about the momentum space structure and physical origin of primordial non-Gaussianity, including signatures associated with excited initial states. The same formalism also shows that the observable is sensitive to the thermal and expansion history preceding BBN, as for a generalized background $a(\eta)\propto\eta^p$, both the damping kernel and the transmission law are modified, while changes in the expansion rate can shift neutrino decoupling and the epoch of photon number freeze out too. A future numerical treatment using full Boltzmann transfer functions could turn relic neutrino measurements into a joint probe of primordial statistics, early Universe microphysics, and non standard pre BBN cosmology. This would end up opening a route to information that is otherwise largely erased from the photon sky.

\appendix

\section{Photon Number Expansion, Kernel Properties, and Thermal Gating}
\label{appendix}
For completeness, we spell out the fixed photon number expansion used above.  Let $m_n=\avg{B^n}_{\hat n}$ and $\avg B=0$.  Through cubic order
\begin{equation}
    \avg{(1+M+B)^4}=1+4M+6M^2+4M^3+6m_2+12Mm_2+4m_3,
\end{equation}
\begin{equation}
    \avg{(1+M+B)^3}=1+3M+3M^2+M^3+3m_2+3Mm_2+m_3.
\end{equation}
Using $(1+x)^{4/3}=1+4x/3+2x^2/9-4x^3/81+\cdots$ then gives
\begin{equation}
 \avg{(1+\Theta)^3}^{4/3}=1+4M+6M^2+4M^3+4m_2+8Mm_2+\frac43m_3,
\end{equation}
This proves \cref{eq:excess}.  Using $n\propto\rho^{3/4}$ at fixed final energy immediately gives \cref{eq:threequarters}. Expanding the cosine in \cref{eq:F-def} and by allowing $t=u^3$ gives the series in \cref{eq:F-def}.  The endpoint expansion $3u^2\e^{-u^3}=3u^2-3u^5+3u^8/2-\cdots$ gives $F(x)=360x^{-6}+\cO(x^{-12})$.  Introducing an Abel regulator $\e^{-\varepsilon x}$ and differentiating
\begin{equation}
 \int_0^\infty\dd x\,\e^{-\varepsilon x}\cos(xu)=\frac{\varepsilon}{\varepsilon^2+u^2}
\end{equation}
proves the moments and a Taylor expansion of ${\Psi(x/r_q)}$ proves \cref{eq:smooth-asymptotic}. For the excited state profile, we may exchange the integrals and use
\begin{equation}
 \int_0^\infty\dd x\,\cos(ux)\frac{\sin(\lambda x)}{\lambda x}=\frac{\pi}{2\lambda}\Theta(\lambda-u).
\end{equation}
\begin{equation}
 \int_0^\infty\dd x\,F(x)\varphi_\lambda(x)=\frac{\pi}{2\lambda}\int_0^\lambda3u^2\e^{-u^3}\dd u=\frac{\pi}{2\lambda}(1-\e^{-\lambda^3}).
\end{equation}
This proves \cref{eq:HT-convolution} and its subsequent.  Inserting $V_\nu(u^3/q)$ gives the general result \cref{eq:HT-transmission-general}. The interplay of the two physical gates is clear when both are instantaneous.  If neutrino decoupling and blackbody freeze out occur at sharp times, we have $V_\nu(y)=\Theta(y-y_{\rm dec})\Theta(y_{\rm bb}-y)$ and $y_{\rm dec}<y_{\rm bb}$. Then
\begin{equation}
     \Feff(x;q)=3\int_{(qy_{\rm dec})^{1/3}}^{(qy_{\rm bb})^{1/3}} \dd u\,u^2\e^{-u^3}\cos(xu)
\end{equation}
\begin{equation}
    \Feff(0;q)=\e^{-qy_{\rm dec}}-\e^{-qy_{\rm bb}}.
\end{equation}
The response is therefore a band pass in damping time, wherein very large $q$ modes dissipate before neutrino decoupling, while very small $q$ modes dissipate after efficient photon production has ended.

\bibliographystyle{unsrt}
\bibliography{bibliography}

\end{document}